 \documentclass[final,3p,times]{elsarticle}

\usepackage{amssymb,amsmath,amsthm, amsfonts, amstext}
\usepackage{footnote}
\usepackage{rotating}
\usepackage{hhline}
\usepackage[normalem]{ulem}
\usepackage{graphicx}
\usepackage{chngcntr}
\usepackage{color}
\usepackage{lineno}

\def \w {\mathrm{w}}
\counterwithout{equation}{section}

\journal{Review of Keynesian Economics}

\nonstopmode

\begin{document}

\begin{frontmatter}

\title{On the Normality of Negative Interest Rates}

\author{Matheus R. Grasselli \footnote{Department of Mathematics and Statistics, McMaster University. Email: grasselli@math.mcmaster.ca. 
Partially supported by a Discovery Grant from the Natural Science and Engineer Research Council of Canada (NSERC).}}
\author{Alexander Lipton \footnote{Connection Science Fellow, Massachusetts Institute of Technology (MIT), and CTO, Sila Money. Email: alexlipt@mit.edu}}

%
\begin{abstract}
We argue that a negative interest rate policy (NIRP) can be an effect tool for macroeconomic stabilization. We first discuss how 
implementing negative rates on reserves held at a central bank does not pose any theoretical difficulty, with a reduction in rates 
operating in exactly the same way when rates are positive or negative, and show that this is compatible with an endogenous money point of view. We then 
propose a simplified stock-flow consistent macroeconomic model where rates are allowed to become arbitrarily negative and present simulation 
evidence for their stabilizing effects. In practice, the existence of physical cash imposes a lower bound for interest rates, which in our view is 
the main reason for the lack of effectiveness of negative interest rates in the countries that adopted them as part of their monetary policy. 
We conclude by discussing alternative ways to overcome this lower bound , in particular the use of central bank digital currencies.
\end{abstract}


\end{frontmatter}



\section{Introduction}
\label{intro}

There is nothing abnormal, in principle, with negative interest rates. Instead of receiving a positive return in an investment, it is entirely possible 
to be charged for holding an asset, especially if all other available assets are deemed to be less safe. 

Nevertheless, positivity used to be considered an essential property in rigorous formulations 
of axiomatic interest rate models \cite{FlesakerHughston96,Rogers97}. The key argument was that, while real interest rates could
be either positive or negative depending on the rate of inflation, negative nominal interest rates would generate an obvious arbitrage opportunity: borrow from a bank
at the prevailing negative rate and invest the funds in zero-interest bearing cash. The argument is robust to extensions of the no-arbitrage principle for 
different lending and borrowing rates \cite{Bergman1995}, in which case it puts a zero lower bound (ZLB) in the deposit rate, as any agent would still prefer to hold funds in cash instead of deposit accounts paying negative interest. 

Guaranteeing that a stochastic process remains positive is not a trivial matter, especially if the source of randomness is a Brownian motion, the favourite choice 
of financial modellers since Louis Bachelier \cite{Bachelier1900}. The landmark Cox-Ingersoll-Ross (CIR) model \cite{CoxIngRos85} has this positivity property 
and has been shown to be a special case of the axiomatic positive interest rate models mentioned above \cite{GrasselliHurd05}. By contrast, the earlier Vasicek model 
\cite{Vasicek1977} leads to normally distributed interest rates, therefore allowing for the occurrence of negative rates with non-zero probability. The underlying Gaussian distribution for the Vasicek model is so appealing from the point of view of analytic tractability, however, that the industry standard consisted of accepting the possibility of negative rates as an ``unrealistic" feature of the model, reinforced by the fact that the probability of such events were very small for typical calibrated parameters 
\cite{BrigoMercurio06}. 

In hindsight, putting up with models that incorporated the possibility of negative interest rates proved to be advantageous to practitioners when this feature became not only realistic but real. In the low rates environment that prevailed after the 2007-2008 financial crisis, instances of negative interest rates became commonplace, not only for short term government bonds, but for a variety of swap rates that permeate the multi-trillion dollars fixed income market. The widespread use of Gaussian models allowed the industry to cope with these ``anomalies" with remarkably little disruption, even if it required ingenious adjustments \cite{Antonov2015}. But if derivative pricing and hedging can be done in practice in the presence of negative interest rates, what does this say about the arbitrage argument mentioned above? The typical answer 
is that holding physical cash creates storage costs that need to be taking into account in order to create an arbitrage, effectively pushing the lower bound into negative territory, the so-called ``physical lower bound". 

More importantly, as negative interest rates seem to be driven by policy initiatives of central banks, are they effective in achieving the desired policy goals? The evidence suggests that the results have been modest, at best \cite{JobstLin2016}. Countries whose central banks have adopted a negative interest rate policy (NIRP) 
do not seem to experience the intended consequences of the policy in any significant way, such as currency depreciation, higher inflation expectations, a robust increase in lending, and 
overall macroeconomic stimulus. In some cases one even observes counterintuitive effects, such as a failure of the traditional transmission mechanism to 
lead to lower rates for loans in Switzerland (see Figure 1, bottom right panel in \cite{JobstLin2016}). Part of the reason for the lack of traction of NIRP is that they have not been pursued beyond very small negative amounts, consistently with the physical lower bound mentioned above. Relatedly, many authors have stressed the existence 
of an ``economic lower bound" for interest rates, below which the adverse effects of lower bank profitability (which is itself a consequence of tighter interest rate 
spreads that arise when negative rates cannot be passed on to depositors because of the physical lower bound) become prevalent \cite{Coeure2016}. 

Some authors \cite{Keen2017,Rochon2017}, however, go further and view NIRP as proof of a fundamental misunderstanding of how banking works on the part of central bankers, in particular with respect to endogenous money. With this, we disagree. As we argue in Section \ref{endogenous}, in our view negative interest rates are entirely consistent with an endogenous money framework of banking. 

In Section \ref{keen_interest} we expand on this argument by showing that a monetary policy rule allowing for sufficiently negative nominal interest rates can be an effective stabilization policy in a stock-flow consistent macroeconomic model. We use an extension of the Keen model \cite{Keen1995} with a consolidate public sector conducting both fiscal and monetary policies. To focus on the effects of 
interest rates, we adopt a simplified formulation of government spending and taxation that ensures that these variables do not alter the qualitative properties of the core dynamical system, which is primarily driven by investment done by the firms sector. As illustration, we provide an example in which rates as negative as -2\% can prevent the explosive debt dynamics typically observed in these types of models. 

Because such levels of negative rates would most certainly violate both the economic and physical lower bounds mentioned above, in Section \ref{crypto} we discuss alternative proposals for 
dealing with these limitations. In particular, because the abolition of physical cash would automatically remove these bounds, we describe a monetary regime in which cash is 
replaced by an electronic and anonymous liability issued by the public sector: a central bank digital currency.

\section{Negative rates and endogenous money}
\label{endogenous}

The central objection to negative interest rates from an endogenous money point view raised in \cite{Rochon2017} is that they were adopted by central banks to deal with the enormous amounts of excess reserves held by banks as a result of the quantitative easing (QE) measures conducted in the aftermath of the financial crisis. As the argument goes, by charging interest on excess reserves, central banks were ``attempting to force banks to lend them out" (page 204). Because ``banks do not 
lend reserves" (page 205), this was an erroneous way to motivate the policy and hence betrays a fundamentally wrong view of how banks operate held by the very people, the central bankers, who are meant to regulate and ultimately support them. This echoes the view expressed in \cite{Keen2017}, where the point that 
``banks can't `lend out reserves' under any circumstance" (page 1) is made even more forcefully through examples of what happens to the total level of 
reserves in the banking system when several illustrative financial transactions take place (such as asset sales between banks). 

Both authors are of course correct in that commercial banks do not lend central bank reserves to their clients. Such reserves are assets held by banks in the form 
of deposits in the central bank and therefore can only be lent to other banks who also have accounts with the central bank. This mechanism of borrowing 
and lending reserves {\em to each other} is in fact the basis of the active Fed Funds market in the United States and similar markets 
for reserves around the world. The problem with the argument in both \cite{Rochon2017} and \cite{Keen2017} is that providing banks with reserves was not 
the motivation for QE and lending reserves out to the general public was not the motivation for NIRP. 

As pointed out for example in \cite{Mehrling2011}, the initial motivation for the unprecedented scale of asset purchases by the 
Federal Reserve\footnote{https://www.federalreserve.gov/newsevents/pressreleases/monetary20081125b.htm} was to stabilize a dysfunctional securities market by playing the role of a ``dealer of last resort". The assets being purchased included all manner of mortgage backed securities and other toxic derivatives for which there were no other buyers in the market, and corresponded to the first round of balance sheet expansion for the Fed in the height of the crisis. Subsequent rounds of QE in the form of purchase of longer-term securities were not intended to provide banks with reserves either, since the initial doses of QE had already left banks awash in excess reserves, but to continue to push up asset prices, consequently reducing their yield and the cost of borrowing across the economy. In other words, while it is certainly possible that some central bankers hold the view that banks are liquidity-constrained and need reserves in order to provide lending, we argue that this was {\em not} the motivation for either the initial rounds of QE, when central banks were purchasing securities in an otherwise frozen market, or its subsequent rounds, when liquidity in the form of systemwide excess reserves was already abundant. 

The argument regarding NIRP is more subtle. We argue that the rationale behind the adoption of negative interest rates should be essentially the same as that for lowering the central bank policy rate from any level that is considered too high for current economic conditions. In other words, a reduction in the policy rate from 2 to 1.75\%, for example, is in no economically essential way different from a reduction from zero to -0.25\%, in particular from the point of view of endogenous money. More specifically, any channels that are expected to help the economy in one case should also be expected to work in the other. 

Take the case for stimulating the economy by increasing lending to the private sector, for example through residential mortgages. Suppose that the prevailing mortgage rate is 4\% when the policy rate is 2\% and suppose that the central bank wants to increase the amount of lending. In normal times the central bank would attempt to achieve this by lowering the policy rate by 0.25\%, for example. This can be implemented in a number of different ways, including a ``symmetric corridor approach", with the policy rate sitting between an upper bound consisting of the discount rate charged on advances from the central bank and a lower bound consisting of interest paid by the central bank on excess reserves, or through so-called ``open market operations", where the central bank buys and sells government assets to change the level of systemwide reserves \cite{Bindseil2004}. No matter the implementation, the end result is that excess reserves end up yielding lower returns to the banks holding them. Profit-seeking banks would then try to replace them with higher yielding assets, such as mortgages. They can try to find more creditworthy customers willing to take up new mortgages at the prevailing 4\% rate, but more likely they would have to offer a lower mortgage rate to attract new customers. For example, a new rate of 3.75\% might generate enough new loans, and consequently new deposits in full accordance with an endogenous money framework, so that previously held {\em excess} reserves become {\em required} reserves (since they are typically a fraction of deposits) and the desired asset substitution is achieved. A complementary action by banks consists in lowering the rate paid on deposits, say from 1\% to 0.75\%, to maintain the same interest rate differential between loans and deposits. 

We should expect the exact same mechanism to work if the prevailing mortgage rate is 2\% and the policy rate is 0\%. Namely, a reduction of the policy rate to -0.25\% 
via interest charged rather than paid on excess reserves makes them unattractive for the banks holding them. Rather then attempt to ``lend them out" to their clients (an accounting impossibility, since the general public does not hold accounts with the central bank and therefore cannot ``borrow" reserves from banks), a profit-seeking 
bank would again try to replace the excess reserves with a higher yielding asset. This can again be done by offering mortgages at a reduced rate of for example 1.75\%, leading to an expansion in loans, and consequently deposits, and corresponding decrease in excess reserves. As before, this can be accompanied by a corresponding reduction in the rate paid on deposits, provided it does not go below the physical lower bound mentioned in Section \ref{intro}.

It is often argued that negative interest on excess reserves functions as cost that banks would {\em naturally} attempt to pass on to customers either through new fees or, perversely, increased loan rates, as was indeed the case in Switzerland \cite{Richter2015}. We remark that such behaviour is as anomalous as banks trying to compensate for the loss of revenue resulting from interest on excess reserves going down from 2 to 1.75\% by {\em increasing} the mortgage rate from 4 to 4.25\% in the example above. While possible, this should not be the expected outcome of a reduction in policy rate, regardless of whether the final rate ends up positive or negative. Indeed, as can be seen in Figure 1 of \cite{JobstLin2016}, the effect of NIRP on lending rates has been exactly as expected: countries that reduced their policy rates into negative territory have experienced lower lending rates in general, with Switzerland being the exception rather than the rule. 

As mentioned in Section \ref{intro}, the empirical evidence of the last few years suggests that macroeconomic effects of NIRP have been 
modest primarily because the rates have been modestly negative. In the next section we present a model in which policy rates are allowed to become arbitrarily negative and investigate its macroeconomic consequences. 

\section{A stock-flow consistent model with negative rates}
\label{keen_interest}

We consider an extension of the Keen model \cite{Keen1995} by adding a government sector with the authority to implement 
monetary policy. A detailed analysis of the original model without the government sector can be found in \cite{GrasselliCostaLima2012}. Because we 
are interested in nominal rather than real interest rates, we consider the version of the Keen model with inflation introduced in \cite{GrasselliNguyenHuu2015}. 
For the extended model, we consider a consolidated public sector conducting both fiscal and monetary polices. On the fiscal side, we assume real government spending and taxation of the form
\begin{align}
G &= g Y \label{spend} \\
T &= t Y \label{tax}
\end{align}
where $Y$ is total real income, and $g$ and $t$ are positive constants. As we will show later (see the discussion preceding equation \eqref{debt_gov}), this 
simplified specification of spending and taxation ensures that fiscal policy does not play any relevant role in the model and allows us to focus exclusively on the effects 
of monetary policy. In addition to taxes, the government finances expenditures by issuing short-term debt $B$ 
satisfying the following budget constraint:
\begin{equation}
\dot B = pG - pT + r_gB,
\end{equation}
where $p$ is a price level and $r_g$ is the short rate of interest paid on government debt. 

The most straightforward way conceptualize short-term government debt in this model is through households having accounts at the central bank, similar to the accounts currently used by commercial banks for the purpose 
of reserve requirements. This is the formulation represented in Table 1, in which case ``issuing short-term debt" simply means crediting the accounts of households, and 
monetary policy consists of setting the interest rate paid on these accounts. Equivalently, we can use a formulation with commercial banks having reserve accounts with the 
central bank. This would split the operation of ``issuing short-term debt" into two steps: the central bank crediting a bank's reserve account, and the bank crediting a household's deposit account. Yet another formulation corresponds to the government actioning short-term bills, with the central bank conducting monetary policy by way of open market operations, that is to say, buying and selling a portion of these bills from banks in exchange of reserves. The important aspect to keep in mind in any of 
these alternative formulations is that $B$ is a short-term liability of the public sector, which therefore has the capacity to set the interest rate $r_g$ by pursuing standard monetary policy implementation mechanisms through its central bank\footnote{Observe that we have in mind the case of a central bank fully backed by a sovereign government and supporting its domestic 
monetary policies. This applies, for example, to the United States, the United Kingdom, and Canada, but does {\em not} apply, for example, to the Euro Zone.}. 

With the previous remarks in mind, we assume that the government adjusts the rate $r_g$ according to the following rule: 
\begin{align}
\label{monetary1}
\dot{\rho} &= \eta_g \left(g_k-\alpha-\beta\right) \\
\label{monetary2}
\dot r_g & =\eta_r(\rho-r_g),
\end{align}
where $\eta_r,\eta_g$ are non-negative constants, $\alpha=\dot a/a$ is the growth rate of productivity and $\beta = \dot{N}/N$ is the growth rate of the labour force, 
and $g_K=\dot{K}/K$ is the growth rate of capital. In other words, the monetary authority adjusts the policy rate $r_g$ to reach a target rate $\rho$, with the target itself taking into account deviations of the real capital growth from the equilibrium growth rate $(\alpha + \beta)$. Crucially, the rule \eqref{monetary1}-\eqref{monetary2} does not imply any lower bound for either $\rho$ not $r_g$, which can 
become significantly negative when the economy is growing far below its full potential. 

For the remainder of the model, we follow closely \cite{GrasselliNguyenHuu2015}. Specifically, we assume that real capital in the economy evolves according to 
\begin{equation}
\dot K = I - \delta K
\end{equation}
where $I$ denotes real investment by firms and $\delta$ is a constant depreciation rate. Capital in turns determines total output through a constant capital-to-output
ratio $\nu = Y/K$. Total nominal output is assumed to be entirely sold and hence equals total nominal income $pY$. Therefore, after paying wages, taxes, interest on loans, and accounting for depreciation, total savings for firms (i.e net profits) are given by 
\begin{equation}
S_f = pY - W - pT - r\Lambda - p\delta K.
\end{equation}
These savings are in turn used by firms to finance investment, with any additional funds being obtained through an increase in loans, that is 
\begin{equation}
\dot\Lambda = p(I-\delta K)-S_f = pI - \Pi_p
\end{equation}
where $\Pi_p = pY - W - pT - r\Lambda$ denotes pre-depreciation profits. We assume that real investment is given by $I = \kappa(\pi)Y$, for an increasing 
function $\kappa(\cdot)$ of the profit share $\pi$ defined as 
\begin{equation}
\pi = \frac{\Pi_p}{pY}=1 - \omega-t- r\ell, 
\end{equation}
where $\omega = W/(pY)$ and $\ell = \Lambda/(pY)$ and $t$ is the constant defined in \eqref{tax}. For the wage-price dynamics we assume that 
\begin{align}
\label{inflation}
\frac{\dot p}{p}& =\eta_p(m\omega-1):=i(\omega) \\
\label{wage}
\frac{\dot{\w}}{\w}& =\Phi(\lambda)+\gamma i(\omega).
\end{align}
In other words, firms set prices in order to achieve a markup $m\geq 1$ over unit labor costs $\mathrm{w}/a$, where $\mathrm{w}$ is the nominal wage rate 
and $a$ is the productivity per worker, whereas workers adjust the nominal wage rate according to an increasing function $\Phi(\cdot)$ of the employment rate 
$\lambda$, while also taking into account the observed inflation rate $i(\omega)=\dot p/p$ via a constant parameter $\gamma$. 

\begin{figure}[!ht]
\centering
\includegraphics[width=0.9\textwidth, height=7cm]{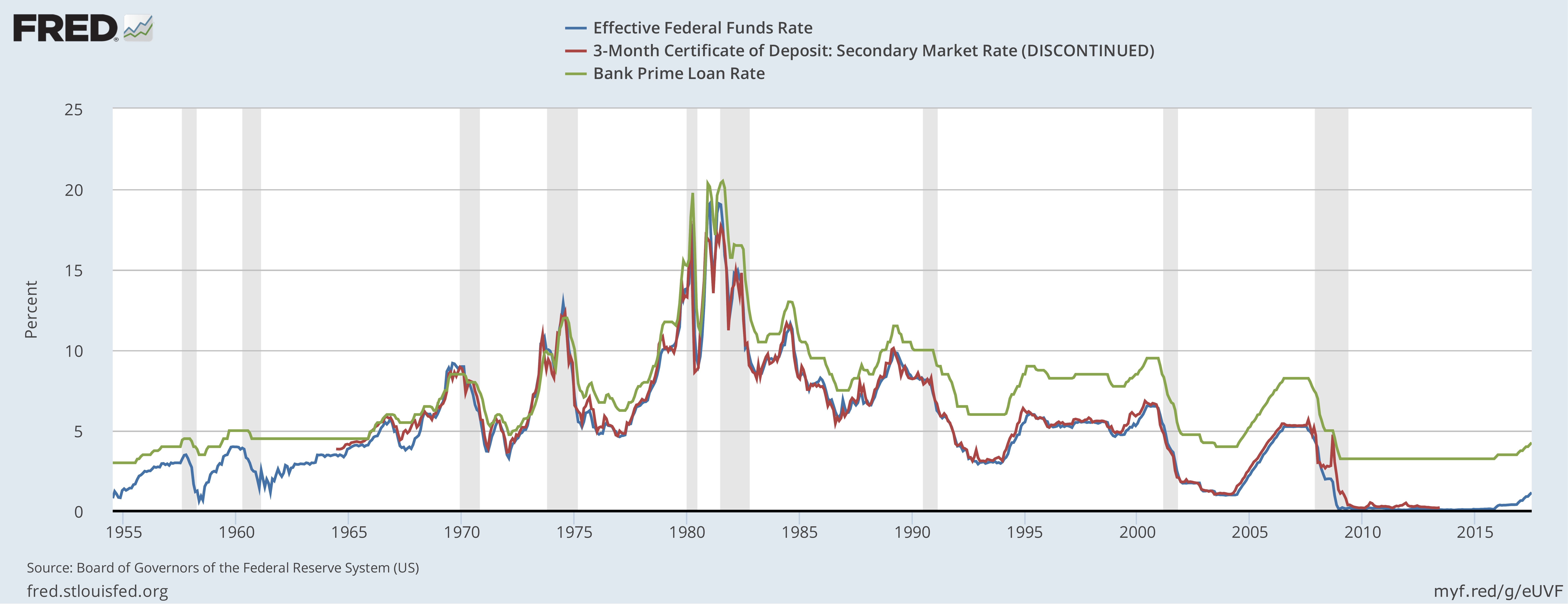}
\caption{The relationship between lending (green), deposit (red) and the policy rate (blue) through time in the United States, supporting our modelling 
choice \eqref{lending_policy}.}
\label{fred_rates}
\end{figure}

We further assume that the interest rate for deposits and loans satisfy  
\begin{equation}
\label{lending_policy}
r_d = r_g \quad \mbox{ and } \quad r = r_g +\delta_r. 
\end{equation}
In other words, the rate $r_d$ offered by banks on deposits is identical to the policy rate $r_g$, whereas the rate charged on loans is the policy rate plus a constant 
spread $\delta_r >0$. While it is certainly possible to incorporate more a complicated interest rate dynamics, this simple rule seems to capture the stylized behaviour of 
observed deposit and lending rates reasonably well, as can be seen in Figure \ref{fred_rates}.

It is then straightforward to see that the model can be described in terms of the following system of ordinary differential equations:
\begin{equation}
\label{keen}
\left\{
\begin{array}{lll}
\dot\omega &= &\omega\left[\Phi(\lambda)-\alpha-(1-\gamma)i(\omega)\right] \\
\dot\lambda  &= &\lambda\left[\frac{\kappa(\pi)}{\nu}-\delta-\alpha-\beta\right]  \\
\dot \ell   &= &\ell \left[ r_g+\delta_r-\frac{\kappa(\pi)}{\nu}+\delta- i(\omega)\right] +\omega+ t+ \kappa(\pi)-1 \\
\dot{\rho} &= &\eta_g \left(\frac{\kappa(\pi)}{\nu}-\delta-\alpha-\beta\right) \\
\dot r_g & = &\eta_r(\rho-r_g)\end{array}
\right.
\end{equation}
where $i(\omega)$ is given by \eqref{inflation} and $\pi=1-\omega-t-(r_g+\delta_r)\ell$. We readily verify that this reduces to the Keen 
model with inflation analyzed in \cite{GrasselliNguyenHuu2015} when $\eta_r=\eta_g = 0$. Moreover, with 
the help of Table \ref{table}, it is easy to see that the model is stock-flow consistent under a variety of specifications of bank behaviour. For example, 
we can assume that banks distribute enough profits to maintain a minimum regulatory capital of the form 
\begin{equation}
\label{regulatory}
X_b=k_r \Lambda,
\end{equation}
where $0<k_r<1$ is a target capital adequacy ratio imposed by the regulators, which we take to the constant. It therefore follows that
\begin{equation}
S_b = \dot X_b = k_r \dot \Lambda = k_r[\kappa(\pi)-1+\omega+(r_g+\delta_r)\ell]pY,
\end{equation}
and 
\begin{equation}
\dot \Delta = \dot \Lambda - \dot X_b = (1-k_r)\dot \Lambda = (1-k_r)[\kappa(\pi)-1+\omega+(r_g+\delta_r)\ell]pY,
\end{equation}
so that both $S_b$ and $\Delta$ can be calculated in terms of the state variables $(\omega,\lambda,\ell,\rho,r_g)$ in \eqref{keen}. 
This in turn completes the model, since the dividends paid by the banking sector are given by
\begin{equation}
\Pi_b=(r_g+\delta_r)\Lambda-r_d\Delta-S_b.
\end{equation}

Importantly, our specification of government spending and taxation in \eqref{spend} and \eqref{tax} ensures that the dynamics for government debt $B$ does 
not affect the main dynamical system \eqref{keen}, since we can verify that the debt ratio $b=B/(pY)$ satisfies 
\begin{equation}
\label{debt_gov}
\dot b = (g-t)+b\left(r_g-i(\omega)-\frac{\kappa(\pi)}{\nu}+\delta\right),
\end{equation}
which can be calculated in terms of the state variables $(\omega,\lambda,\ell,\rho,r_g)$ in \eqref{keen}. We made this choice deliberately so that any stabilizing effect that the government might have on \eqref{keen} comes exclusively from the monetary 
policy rule adopted in \eqref{monetary1}-\eqref{monetary2}. That is not to say that we think there is no stabilizing role for fiscal policy in this class of models (as can 
be seen for example in \cite{CostaLimaGrasselliWangWu2014}), but rather that we decided to 
isolate the effects of monetary policy instead. Similarly, we emphasize that the model in \eqref{keen} can be generalized in a number of ways, including 
the stochastic extensions and other improvements to the base model provided in \cite{Lipton2016b}.

\begin{table*}[t]
	\centering
			\begin{tabular}{|l|c|cc|c|c|c|}
				\hline
				& Households & \multicolumn{2}{|c|}{Firms} & Banks &  Public & Sum  \\
				\hline 
				{\bf Balance Sheet} &  & &  & & &  \\
				Capital stock &  & \multicolumn{2}{|c|}{$+pK$} & &   & $+pK$ \\
				Deposits & $+\Delta$ & \multicolumn{2}{|c|}{} &   $-\Delta$& & 0  \\
				Loans &  & \multicolumn{2}{|c|}{$-\Lambda$} &  $+\Lambda$ & &  0 \\
				Bills & {$+B$} &  & & & $-B$ & 0\\
				\hline
				Sum (net worth) & $X_h$ & \multicolumn{2}{|c|}{$X_f$}  & $X_b$  & $X_g$ & $pK$ \\
				\hline 
				\hline
				{\bf Transactions} & &  current & capital & &   &\\
				Consumption  & $-pC$ & $+pC$ & &  &  &  0 \\
				Gov Spending & & $+pG$ & & & $-pG$ & 0 \\
				Capital Investment  & & $+pI$ & $-pI$ & &  & 0  \\
				Accounting memo [GDP] & & [$pY$]  & & & & \\
				Wages & $+W$ & $-W$ & &  & & 0 \\
				Taxes & & $-pT$ & & & $+pT$ &  0\\
				Depreciation & &$-p\delta K$ &$+p\delta K$ & & & 0 \\
				Interest on deposits  & $+r_{d}\Delta$ &  & &   $-r_{d}\Delta$ &  &   0 \\
				Interest on loans  &  & $-r\Lambda$ &  & $+r\Lambda$ & &   0 \\
				Interest on Bills &  $+r_gB$ &   & & & $-r_gB$  & 0 \\
				Dividends & $+\Pi_b$ & & & $-\Pi_b$ & &     0\\
				\hline
				Financial Balances & $S_h$ & $S_f$ & $-p(I-\delta K)$ & $S_b$ & $S_g$ & 0 \\
				\hline
				\hline
				{\bf Flow of Funds} & &  &    & &   &\\
				Change in Capital Stock & & \multicolumn{2}{|c|}{$+p(I-\delta K)$}& & & $+p(I-\delta K)$\\
				Change in Deposits & $+\dot{\Delta}$& \multicolumn{2}{|c|}{}&  $-\dot{\Delta}$& &   0  \\
				Change in Loans & & \multicolumn{2}{|c|}{$-\dot{\Lambda}$}   &  $+\dot{\Lambda}$ &  &  0 \\
				Change in  Bills & $+\dot{B}$ &   &  & & $-\dot{B}$ & 0 \\
				\hline
				Column sum & $S_h$ &  \multicolumn{2}{|c|}{$S_f$}  &   $S_b$ & $S_g$ &  $p(I-\delta K)$\\ \hline
				Change in net worth & $S_h$ &\multicolumn{2}{|c|}{$S_f+\dot{p}K$}   &$S_b$ & $S_g$ &  $\dot pK +p\dot K$\\
				\hline
			\end{tabular}
		\caption{Balance sheet and transactions flows.}
		\label{table}
\end{table*}%

It is straightforward to see that \eqref{keen} admits an interior equilibrium of the form 
\begin{eqnarray}
\label{eq:omega^1_in} \overline\omega_1 &=& 1 - \overline\pi_1 -t - (\overline r_0+\delta_r) \overline \ell_1 \\
\label{eq:lambda^1_in} \overline\lambda_1 &=& \Phi^{-1}[\alpha + (1-\gamma)i(\overline\omega_1)] \\
\label{eq:ell^1_in} \overline \ell_1 &=& \frac{\kappa(\overline\pi_1) - \overline\pi_1}{\alpha+\beta+i(\overline\omega_1)} \\
\label{eq:r^1_in} \overline r_g &=& \overline \rho
\end{eqnarray}
where $\overline \pi_1:= \kappa^{-1}(\nu(\alpha+\beta+\delta))$ is an equilibrium profit rate, with the corresponding equilibrium government debt ratio is 
then given by 
\begin{equation}
\overline b_1 = \frac{g-t}{i(\overline\omega_1)+\alpha+\beta-\bar r_g}. 
\end{equation}

An analysis of other possible equilibria of \eqref{keen} and their stability properties can be conducted very similarly to \cite{GrasselliNguyenHuu2015} and 
is omitted here in the interest of brevity. We demonstrate the stabilization effects of the monetary policy rule \eqref{monetary1}-\eqref{monetary2} by means of a 
few illustrative examples instead. We adopt a 
a Philips curve and investment function of the form
\begin{align}
\Phi(\lambda)&= a+\frac{b}{(1-\lambda)^2} \\
\kappa(\pi) &= c+\exp(d+e\pi)
\end{align}
and, unless otherwise stated, we use the parameters listed in Table \ref{parameters}.

\begin{table}[!ht]
\begin{center}
  \begin{tabular}{c|l|l}
 Symbol & Value & Description \\ \hline
$a$ & -0.0401 & constant term in Phillips curve \\
$b$ & 0.0001 & coefficient in Phillips curve \\
$c$ & -0.0065 & constant term in investment function \\
$d$ & -5 & affine term in exponent of investment function \\
$e$ & 20 & coefficient in exponent of investment function  \\
$k_r$ & 0.1 & capital ratio for banks \\
$m$ & 1.3 & markup factor \\
$\alpha$ & 0.025 & growth rate in productivity  \\
$\beta$ & 0.02 & growth rate in labour force  \\
$\gamma$ & 0.8 & money illusion coefficient  \\
$\delta$ & 0.03 & depreciation rate \\
$\eta$ & 0.35 & inflation relaxation parameter \\
$\nu $ & 3 & capital to output ratio \\
\hline
  \end{tabular}
 \caption{Baseline parameter values}
 \label{parameters}
\end{center} 
\end{table}

\begin{figure}[!ht]
\centering
\includegraphics[width=0.9\textwidth, height=9cm]{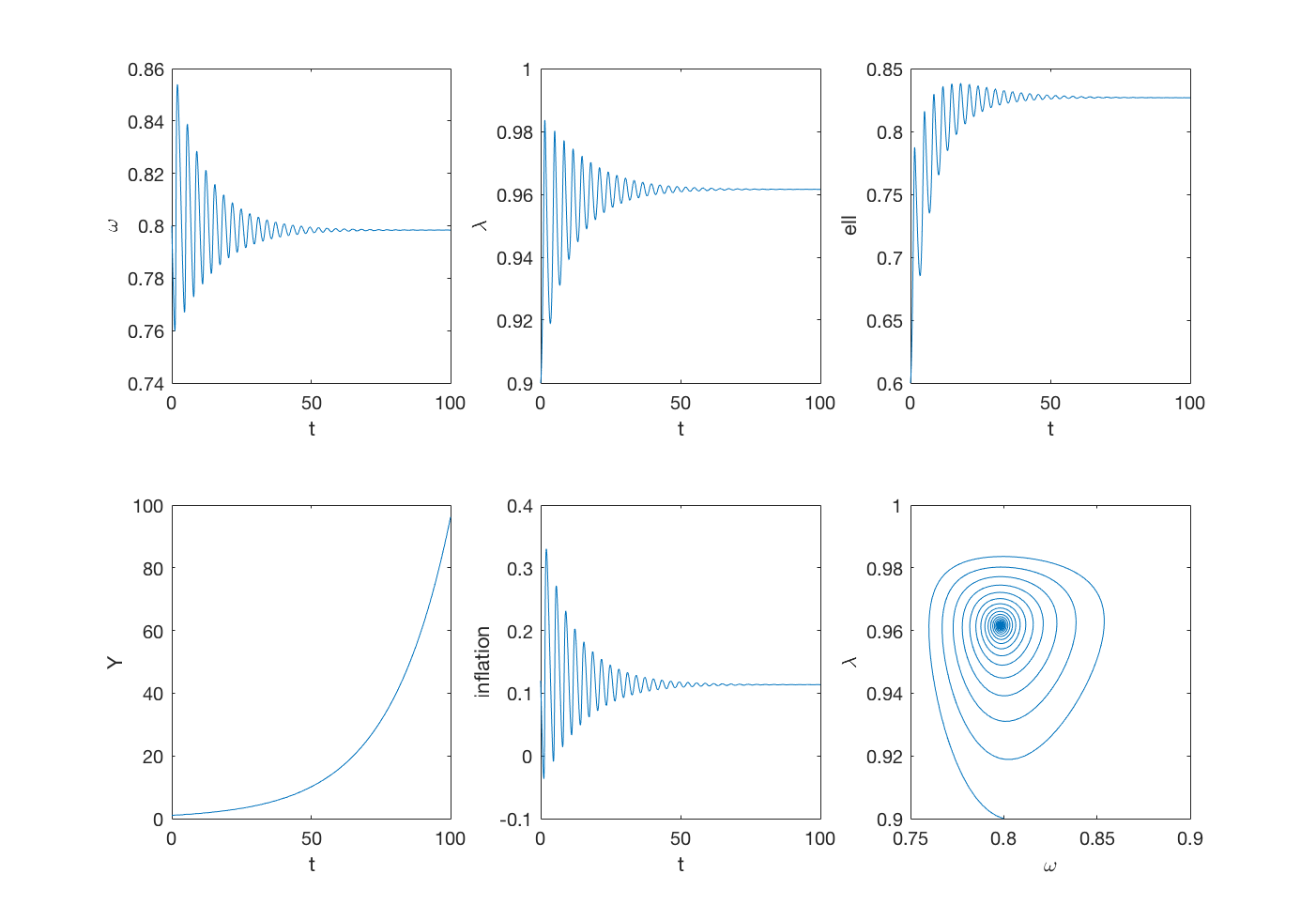}
\caption{Solution of the Keen model \eqref{keen} without monetary policy. The parameter values in addition to those in Table \ref{parameters} 
are $g=t=\eta_r=\eta_g=0$ and $r=0.03$. With initial conditions $\omega_0=0.8$, $\lambda_0=0.9$ and $\ell_0=0.6$ (moderate level of private debt), we observe convergence to an interior equilibrium.}
\label{keen1}
\end{figure}

We begin with an example of the Keen model in the absence of any active monetary policy (i.e with $\eta_r=\eta_g=0$ and 
$g=t=0$ for simplicity) shown in Figure \ref{keen1}. We can see that the benign initial conditions, namely high wage share and employment rate combined 
with moderate levels of private debt, ensure that the model converges to an equilibrium with finite private debt ratio, positive wage share, and 
positive employment rates. As we can see in the figure, the level of real output grows steadily whereas the rate of inflation converges to a moderate positive value. 

\begin{figure}[!ht]
\centering
\includegraphics[width=0.9\textwidth, height=9cm]{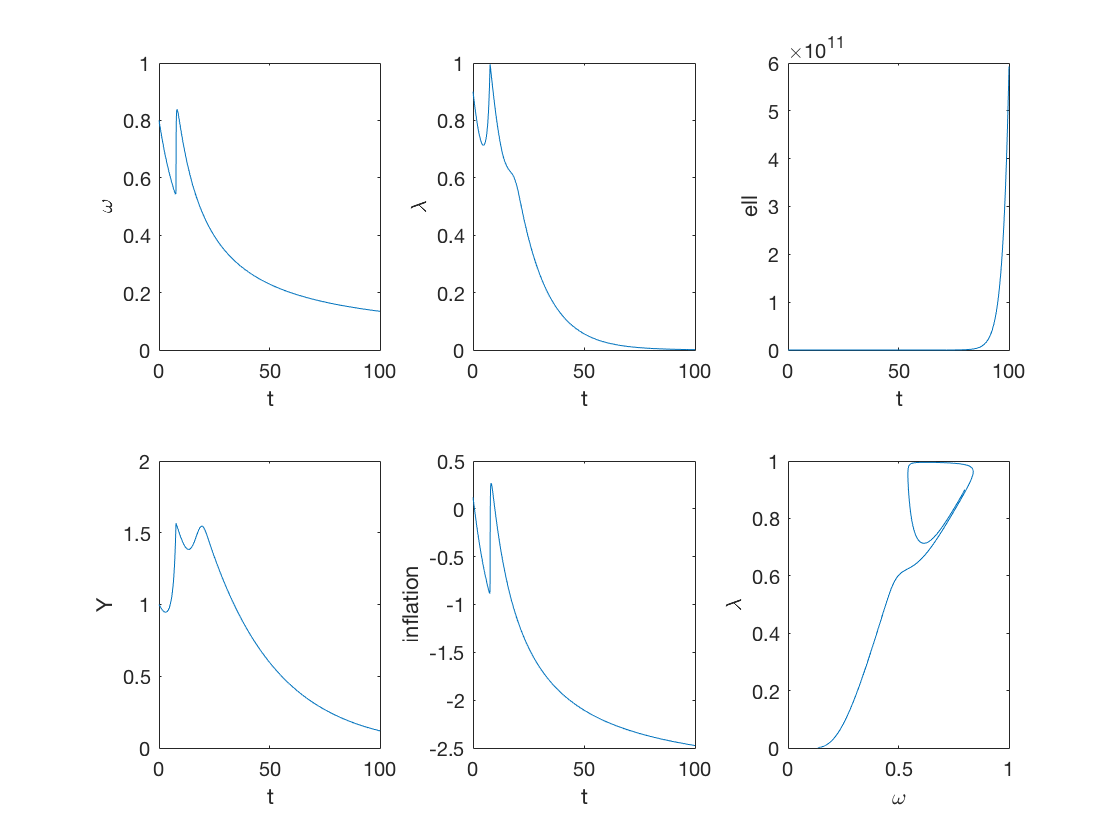}
\caption{Solution of the Keen model \eqref{keen} without monetary policy. The parameter values in addition to those in Table \ref{parameters} 
are $g=t=\eta_r=\eta_g=0$ and $r=0.03$. With initial conditions $\omega_0=0.8$, $\lambda_0=0.9$ and $\ell_0=6$ (high level of private debt), we observe explosive private debt, collapsing output and asymptotic deflation.}
\label{keen2}
\end{figure}

We follow this by an example of the exact same model but starting from a much higher level of private debt. As we can see 
in Figure \ref{keen2} this leads to an explosive trajectory where the private debt ratio grows unboundedly and the wage share and employment rate both collapse to zero. 
Consequently, after an initial period of growth, output decreases monotonically and inflation converges to a negative asymptotic value. This is exactly the scenario 
that an aggressive monetary policy aims to prevent. 

\begin{figure}[!ht]
\centering
\includegraphics[width=0.9\textwidth, height=9cm]{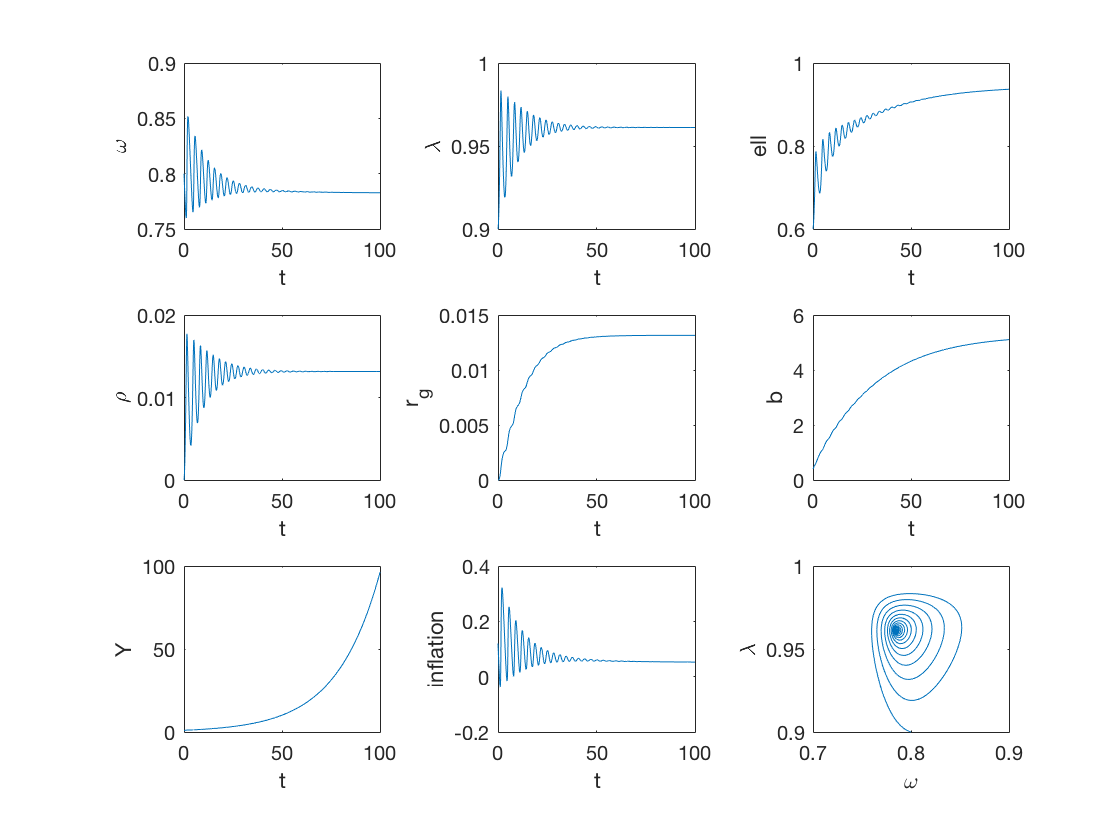}
\caption{Solution of the Keen model \eqref{keen} with monetary policy. The parameter values in addition to those in Table \ref{parameters} 
are $g=0.2$, $t=0$, $\delta_r=0.03$, $\eta_r=0.1$ and $\eta_g=0.2$. With initial conditions $\omega_0=0.8$, $\lambda_0=0.9$, 
$\ell_0=0.6$ (moderate level of private debt), $\rho=0$, $r_g=0$, $b=0.4$, we observe convergence to an interior equilibrium.}
\label{keen3}
\end{figure}

In the next three examples we explore the Keen model \eqref{keen} with full monetary intervention by the government. Figure \ref{keen3} shows the effect 
of monetary policy in the case of the same benign initial conditions as in Figure \ref{keen1}. We see that the target rate $\rho$ in this case oscillates 
around small but positive values before converging to an asymptotic value slightly above 1\%, leading the policy rate $r_g$ to increase from zero to the same 
asymptotic value. In such favourable economic scenario, monetary policy intervention has 
very little effect on the qualitative properties of the model, which converges to essentially the same equilibrium as in Figure \ref{keen1}, 
with a slightly higher private debt ratio (due to the fact that the lending rate converges to a value slightly higher than 4\%, compared to $r=0.03$ in Figure \ref{keen1}, 
growing output and small but positive inflation. The government debt ratio converges to a relatively high but sustainable value slightly above 4, which in practice can be 
made smaller by adjusting the difference $g-t$ (we chose the unrealistic value $t=0$ to make the example as comparable as possible to that of Figure \ref{keen1}). 

\begin{figure}[!ht]
\centering
\includegraphics[width=0.9\textwidth, height=9cm]{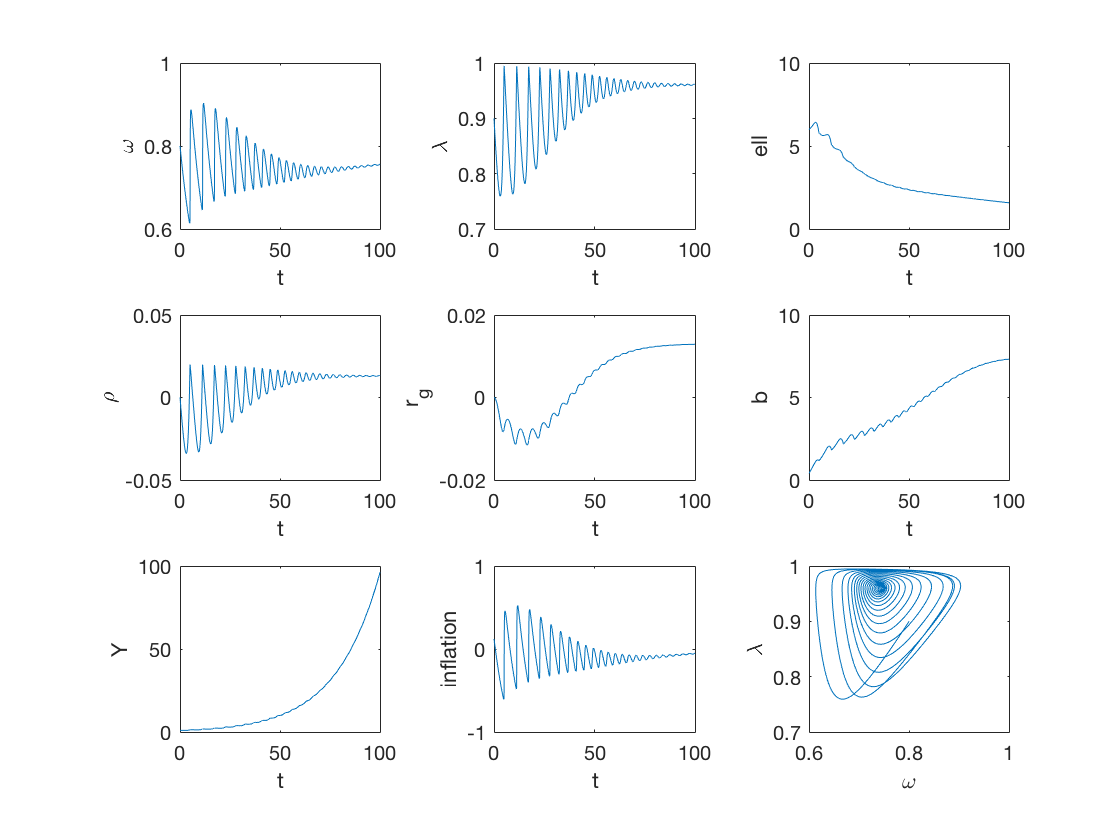}
\caption{Solution of the Keen model \eqref{keen} with monetary policy. The parameter values in addition to those in Table \ref{parameters} 
are $g=0.2$, $t=0$, $\delta_r=0.03$, $\eta_r=0.1$ and $\eta_g=0.2$. With initial conditions $\omega_0=0.8$, $\lambda_0=0.9$, 
$\ell_0=6$ (high level of private debt), $\rho=0$, $r_g=0$, $b=0.4$, we still observe convergence to an interior equilibrium.}
\label{keen4}
\end{figure}

\begin{figure}[!ht]
\centering
\includegraphics[width=0.9\textwidth, height=9cm]{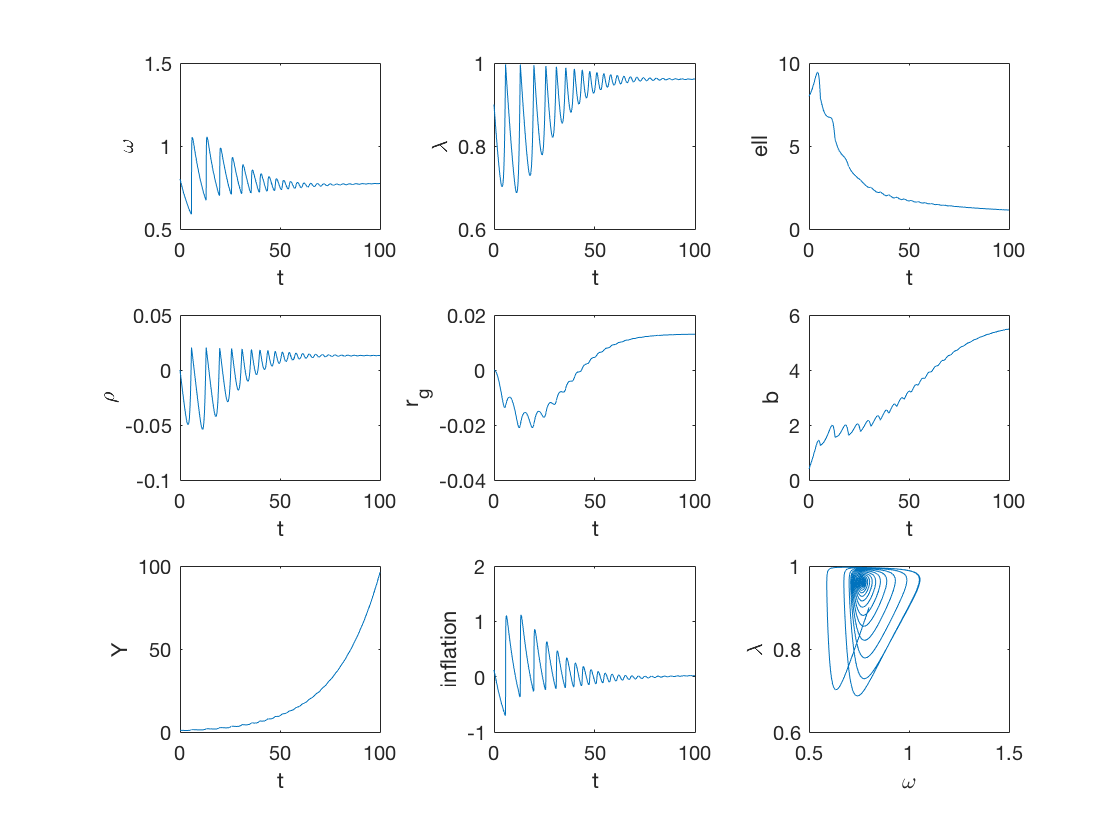}
\caption{Solution of the Keen model \eqref{keen} with monetary policy. The parameter values in addition to those in Table \ref{parameters} 
are $g=0.2$, $t=0$, $\delta_r=0.03$, $\eta_r=0.1$ and $\eta_g=0.2$. With initial conditions $\omega_0=0.8$, $\lambda_0=0.9$, 
$\ell_0=8$ (even higher level of private debt), $\rho=0$, $r_g=0$, $b=0.4$, we still observe convergence to an interior equilibrium, but with 
the policy rate $r_g$ becoming significantly negative for a while.}
\label{keen5}
\end{figure}

Figure \ref{keen4} is where we begin to see the effect of negative interest rates in earnest. Using the same unfavourable initial conditions as in Figure \ref{keen2}, namely a high initial level of private debt, we see that letting the policy rate dip as low as $-1\%$ has the beneficial 
effect of bringing down the private debt ratio, maintaining output growth, and preventing asymptotic deflation. Observe that the policy rate $r_g$ does
not need to remain negative indefinitely, but on the contrary moves into positive territory once the growth rate of output recovers sufficiently. 
In Figure \ref{keen5} we push the initial level of private 
debt even higher and observe the same results, but at the price of having the policy rate $r_g$ dip close to $-2\%$ before returning to 
positive values. 

The last example is arguably extreme, but it illustrates that for a NIRP to be truly stabilizing, 
nominal interest rates need to be allowed to become sufficiently negative. Because this is likely to violate the typical values for the 
physical and economic lower bounds mentioned in Section \ref{intro}, one is lead to explore ways to make these bonds even lower or not 
applicable altogether, as we describe in the next section. 

\section{Central bank digital currency and negative rates}
\label{crypto}

One of the most common arguments traditionally used to justify the zero
lower bound (ZLB) for interest rates as self-evident is the suggestion that
depositors can escape negative interest by shifting their deposits into
cash. Putting aside that fact that this is a very remote possibility in a
modern developed economy for a simple reason that the amount of cash
constitutes but a small fraction of the aggregate money supply\footnote{As of April 2018, notes in circulation made 
up 8\% of M1, 5\% of M2, and 3\% of M3 in Canada. Source: Statistics Canada, Table  10-10-0116-01}, it is worth considering what kind of the lower bound
can be achieved in the presence of cash. In general, holding significant
amounts of cash is relatively difficult, especially if very large
denominated banknotes are not available. Storing, protecting, and insuring
cash results in a negative carry of the order of 30 bp (30 basis points, or 0.3\%). The expenses associated with
transacting in cash (for example, the time spent counting notes) adds another 20 bp, say, so that the total is 50 bp.
Thus, cash-related physical lower bound (PLB) mentioned in Section \ref{intro} is about -0.5\%. 

In order to have negative rates lower than this PLB, as the simulations of Section \ref{keen_interest} suggest might be necessary for 
a NIRP to have significant macroeconomic effects, three possibilities can be contemplated: (A) taxing
cash; (B) breaking the parity between bank deposits and cash; (C) abolishing cash altogether. 

The idea of taxing cash is not new. For
example, it was practiced in Europe in the High Middle Ages it in the form
of demurrage. In simple terms, Medieval treasuries periodically recalled
gold and silver coins, re-minted them and returned some of the new coins to
the original owners, while retaining some of the coins at the treasury.%
\footnote{%
This practice should not be confused with a more common practice of currency debasement, whereby the quantity of gold, silver, copper, or nickel in a coin is reduced.}
In effect, demurrage was a tax on monetary wealth, which required a massive
apparatus of coercion to implement it efficiently. More recently, such tax
was advocated by S. Gesell \cite{Gesell1958} and I. Fisher \cite{Fisher1945} in the
form of the so-called stamp script. It is clear that in the presence of such
a tax, its magnitude plus the size of frictions related to holding cash 
determine the ultimate PLB. However, taxing cash might be too unpalatable
for modern sensitivities. 

The second alternative consists of having a market-determined exchange rate 
between bank deposits and cash. This possibility is explored in detailed in \cite{Goodfriend2016}, where it is observed 
that negative interest rates would theoretically drive the deposit price of cash above par in the same way 
that the price between two different currencies react to interest rate differentials. While intriguing, such floating deposit-to-cash 
exchange rates would like suffer from the same drawbacks of floating exchange rates for currencies, 
including periods of high volatility.  

In this section we explore the idea of abolishing cash altogether as a realistic option. 

We start with the observation that cash, in the form of coins and notes in circulation, is a liability of 
the central bank that is held as an asset by the general public. By contrast, as mentioned in Section \ref{endogenous}, reserve accounts are liabilities 
of the central bank that are currently held as assets by commercial banks only. Because reserve accounts are already electronic, it is 
therefore tempting to argue that an obvious way to abolish cash is to allow the general public to also have similar with the central bank. Such accounts, however, would fail to have one key property of physical cash: anonymity. Much in the same way that a bank knows the identity of all its depositors, as well as the amounts held in each deposit account, the central bank would know the owner and the amount held in each of these general-public reserve accounts. 

The real challenge consists in replacing physical cash, which corresponds to a sizeable component of the liabilities of any central bank\footnote{As of April 2018, notes in circulation (about CAD\$ 83 billion) correspond to 75\% of the liabilities of the Bank of Canada. Source: Statistics Canada.  Table  10-10-0108-01.} with an alternative that is both electronic and anonymous. Until recently, this was not technically feasible. However, achievements of the
modern cryptography combined with great increases in computer power make it
possible to abolish physical cash and replace it with central bank issued
digital cash (CBDC) \cite{Lipton2016}. It is not a coincidence that a ``great war" on paper money has
recently started in earnest \cite{Rogoff2016}. Both academics and
practitioners are currently engaged in heated debates about the desirability
of survival of physical cash. These debates are fuelled by the introduction
of highly successful cryptocurrencies such as Bitcoin and Ethereum on the
one hand, and perceived necessity to introduce significant negative interest
rates on the other.

The feat of engineering required to issue CBDC cannot be overestimated. It is unlikely to be achieved by using Bitcoin-like technology a la S. Nakamoto 
\cite{Nakamoto2008}. A creative reuse of D. ChaumÕs ideas \cite{Chaum1990} describing a digital currency based on cryptographic protocols seems to be more promising. Let us briefly describe the corresponding approaches.

Bitcoin relies on distributed ledgers and blockchains (see \cite{Lipton2018} for a more extensive introduction). In general, ledgers come in several flavours in increasing order of complexity from traditional centralized ledgers to unpermissioned public ledgers. Bitcoin relies on the most complex choice by using an unpermissioned public ledger with proof of work (POW) verification mechanism to solve the double-spend problem (see \cite{Lipton2018}). The verification is done by miners, who are rewarded for their work with new bitcoins, but with the reward rate halved at regular intervals, so that the total number of bitcoins converges to 21 million. The complexity of the bitcoin ecosystem 
leads to a very low number of transactions per second (TPS)\footnote{For Bitcoin, TPS=7. By comparison, for Ethereum TPS=20, for Paypal TPS=200, and for Visa TPS=2000-20000.} at enormous real costs\footnote{Bitcoin consumes as much electricity as the country of Croatia}. It is important to realize, however, that these limitations were introduced by design in order to achieve a system with an asymptotically constant number of coins and no central authority. Neither of these features is necessary for CBDC. 

Instead, one can set up a system where CBDC is issued in a centralized (directly by the central bank), or a semi-centralized (by narrow banks doing central bank bidding) fashion. The double-spend problem is resolved by using a single-usage unique identifier that is issued by the central bank or its authorized representatives. In order to achieve the required speed and lower computational burdens, even if the identifier itself is centrally issued, it should be verified by a set of validators (or notaries), potentially with each of them knowing only a part of the overall number, so that the system cannot be abused. The privacy can be achieved by using ChaumÕs blind signature protocol. Alternatively, designated (rather than anonymous) validators can perform the role of Bitcoin miners but without prolific and unsustainable electricity spending. 

With cash out of the picture, interest rates could be set as negative as
determined by the current economic insight of central bankers, for example guided by 
models such as the one discussed in Section \ref{keen_interest}. In addition to being instrumental for implementing negative interest rate
policies, CBDC opens additional exciting possibilities. In particular, CBDC
makes the execution of the celebrated Chicago Plan of 1933 \cite{KumhofBenes2012} for
introducing narrow or full-reserve banks, which hold as much central bank
cash as they have deposits, a possibility within reach\footnote{%
The idea of a narrow bank was originally introduced by D. Ricardo in \cite{Ricardo1824}
and further expanded by F. Soddy in \cite{Soddy1926}.}.
Some authors (see \cite{BarrdearKumhof2016} for example), suggest that once CBDC becomes
prevalent, both firms and ordinary consumers would be able to use it to increase their direct holdings of 
central banks liabilities. The aggregate deposits at commercial banks would 
naturally decrease, potentially by a wide margin. As a result, commercial
banks would lose their central position in the economy and become akin to
credit mutual funds. In order to retain their ability to create money ``out
of thin air", banks might contemplate issuing their own money (naturally
paying positive interest), thus creating a somewhat paradoxical situation
when narrow banking and free banking coexist side by side. This line of
reasoning omits one important point, namely, the fact that, in addition to
many other roles they play, commercial banks are gatekeepers of the
financial system as a whole and are expected to know their customers and
combat money laundering. Besides, relying on central banks ability to calculate the
``correct" yield curve for a given set of economic circumstances, rather than
letting this curve to emerge as a result of free market forces, seems to be
na\"{i}ve. Given the sheer complexity of these tasks, it is
clear that central banks are not able to solve them reliably, so a demise
of banking system caused by the introduction of CBDC is unlikely to happen. However, the emergence of some 
forms of narrow banking is to be expected in
the near future. 

We mention in passing that CBDC could potentially make criminal activities,
such as drug trafficking and tax evasion, more difficult, thus greatly
benefiting society at large. In addition, it would facilitate transactional
banking, and make it more inclusive by allowing the unbanked to participate
in the digital economy. At the same time, in the negative interest rate
world, anyone who has to rely on fixed income instruments for fulfilling
their investment objectives, have to find some other means to achieving them.

Apart from technical considerations, another practical roadblock to the implementation of negative rates on 
all money concerns the issues of fairness and equity, which parallel those associated with sales taxes that are imposed without 
exclusions for necessities. Such concerns help to generate political opposition to the elimination of a stable unit of account for small transactions and 
bank accounts of households with low net worth. To alleviate this problem, governments might consider allowable exceptions, for example in the 
form of an exemption from negative rates on bank reserves in proportion to the amount of deposits kept in small bank accounts.

\section{Concluding remarks}

The topic of negative interest rates elicit strong reactions from nearly every corner of the ideological spectrum. Traditional Keynesian economists
tend to view them as ``monetary wizardry" exacerbating the ``folly of relying on monetary policy alone to rescue economies from depressed conditions"  \cite{Skidelsky2016}, whereas Austrian economists see them as as example of ``radical monetary policy" that ``will only ensure an ever greater misalignment between output and demand" \cite{Hollenbeck2016}. In this article, we have taken the point of view that there is nothing intrinsically abnormal about negative interest rates. 

To begin with, we have argued in Section \ref{endogenous} that the introduction of negative interest rates by central banks around the world does not betray a fundamentally flawed understanding of banking in general and endogenous money in particular. On the contrary, negative interest rates are entirely compatible 
with the fact that banks can offer loans, and create deposits in the process, without being constrained by reserves. Moreover, the transmission mechanisms that are expected to work at positive rates, for example an increase in lending to the private sector associated with a reduction in the policy rate, should in theory operate in essentially the same way when rates move into negative territory. 

In Section \ref{keen_interest} we elaborate on the theoretical effectiveness of negative interest rates by constructing a stock-flow consistent macroeconomic model in which the policy rate has a significant stabilizing role, provided it is allowed to become 
sufficiently negative during periods of slow economic growth. The intuition behind the model is that lower interest rates improve the financial position of 
firms, so that they can invest more and increase aggregate demand. As with any variant of the Keen model without an independent saving function for 
households, our model ignores the effects of consumption on aggregate demand, and a more complete model would have to take into account the decrease in 
consumption caused by lower interest revenue. Although this is certainly a deficiency in the model presented here, it is possible to argue that 
households with significant interest rate revenue tend also to be rich, and therefore have a lower propensity to spend to begin with, so that negative interest rates 
would have a smaller effect on consumption than on investment. In any case, the purpose of the model in Section 3 was to show that if lower interest rates have a 
stabilizing effect, than this remains true when they are allowed to be negative. 

In practice, of course, central banks in places like Japan, Sweden, Switzerland, and the Eurozone have only adopted modestly negative interest rate policies, typically above -1\%, with predictably modest results. It is generally argued that the reason for this is that the possibility of holding zero-interest bearing paper cash prevents 
deposit rates from becoming arbitrarily negative, forcing them to remain above a so-called physical lower bound. In Section \ref{crypto} we discuss ways to move 
beyond this lower bound, in particular the replacement of paper cash by purely digital currencies issued by central banks. 

Preventing and fighting economic recessions are major tasks that ought to be tackled by a multitude of tools and approaches. Dismissing negative interest rates from 
the outset as either ineffective or absurd seems to be counter productive. On the other hand, timid experimentation with NIRP, limited by either physical or 
psychological barriers associated with paper cash, only lead to doubts and confusion about its goals and effectiveness.

\end{document}